\documentclass[fleqn,10pt]{wlscirep}
\usepackage[utf8]{inputenc}
\usepackage[T1]{fontenc}
\usepackage{bm}
\usepackage[utf8]{inputenc}
\usepackage{amsmath}
\usepackage{multirow}
\usepackage{graphicx}
\usepackage{authblk}
\usepackage{floatrow}
\usepackage{changepage} 

\begin{document}
\flushbottom

\title{Recreational Mobility Prior and During the COVID-19 Pandemic}

\author[1]{Zahra Ghadiri}
\author[2]{Afra Mashhadi}
\author[3]{Marc Timme}
\author[1,3 *]{Fakhteh Ghanbarnejad}

\affil[1]{Department of Physics, Sharif University of Technology, Tehran, Iran}
\affil[2]{Computing and Software Systems, University of Washington, Bothell, WA, USA}
\affil[3]{Chair for Network Dynamics, Institute for Theoretical Physics and Center for Advancing Electronics Dresden (cfaed), Technical University of Dresden, 01062 Dresden, Germany}
\affil[*]{fakhteh.ghanbarnejad@gmail.com}

\begin{abstract}
    The COVID-19 pandemic and the resulting economic recession  negatively affected many people’s physical, social, and psychological health and has been shown to change population-level mobility, but little attention has been given to park visitations as an indicator.  Estimating the frequency of  park visitations  from  aggregated mobility data of all the parks in   Washington State (USA), we study  trends in park use one year prior to and two years during the COVID-19 pandemic. Our findings indicate that the gravity model is a robust  model for the  park visitation behavior in different spatial resolutions of city level and state level and different socio-economical classes.    Incorporating network structure, our detailed analysis highlights that  high-income level residents changed their recreational behavior by visiting their local parks more and a broader recreational options outside of their local census area; whereas the low-income residents changed their visitation behavior by reducing their recreational choices. 
    
\end{abstract}

\maketitle
\section{Introduction}

Parks and recreational systems serve as vital connectors, linking both residents and visitors to the beauty of nature, offering opportunities for physical and mental well-being ~\cite{russell2013humans,soga2021room,remme2021ecosystem,keniger2013benefits}, and fostering social interaction and community cohesion~\cite{keeler2019social,stodolska2015recreation,seattlerec}.  Having information about  park visitation and usage is an important first step for policymakers to be able to address and plan for fair and equitable access to parks~\cite{seattlerec} especially during significant societal disruptions such as those observed during the COVID-19 pandemic.   During the COVID-19 pandemic, many people shifted to remote working and different socio-economic classes exhibited disparities in their mobility choices as well as leisure time and affordances to access parks and recreational sites. At the population level, some evidence indicated that park visitation increased in 2020~\cite{geng2021impacts,lu2021escaping,rice2020changes}. However several recent studies on the impact of the pandemic on park visitation have concluded different observations. A global analysis of park visitation patterns using data collected primarily from Google’s Community Mobility Reports~\cite{google} showed that park visitation has increased since February 2020, compared to visitor numbers prior to the COVID-19 pandemic \cite{geng2021impacts}.  Other studies, such as~\cite{jay2021effects} showed a 14\% decline in urban park visitation during the pandemic, based on a study of US urban parks using data from SafeGraph~\cite{safegraph}. In~\cite{geng2021impacts} authors conducted a longitudinal study of  Instagram posts from 100K users (1,185 green spaces) in four Asian cities and showed a 5.3\% surge during the pandemic.  Suse et al.\cite{suse2021effects} also used Instagram data to show that there are contrasting observations across 4 US metropolitan cities and argued more contextual information  is needed to explain the differences. Despite its importance, little attention has been given to understanding and modeling the  {statistical and structural patterns} of park visitations and how these patterns have been changed in regards to seeking nature and recreational opportunities during large societal phenomena such as the COVID-19 pandemic.

In a broader domain of literature,  human mobility has been  studied in the past decades~\cite{barbosa2018human} from various perspectives.   Existing large-scale mobility studies have applied  state-of-the-art models such as the gravity model~\cite{zipf1946p,erlander1990gravity,ribeiro2023mathematical,altmann2020spatial}, the radiation model~\cite{simini2012universal}  and others~\cite{noulas2012tale,stouffer1940intervening,song2010modelling}  to  seek  understanding of city structure~\cite{anas1998urban,bassolas2019hierarchical,barlacchi2015multi,louail2015uncovering,henderson2004handbook},  epidemics~\cite{chang2010epluribus,eubank2004modelling,wesolowski2012quantifying}, and visitation patterns~\cite{schlapfer2021universal}.

In order to improve our understanding of park visitation, here we   seek to model park visitations into the underlying statistical and structural patterns of travel distance over a period of two years. To do so, we borrow tenets from mobility literature and apply the state-of-the-art mobility model based on the  {Gravity Law} that has successfully modeled mobility by concentrating on the spatial dependence of population flows. The gravity model has been  widely applied in the past with great success in modeling the mobility behavior of people across the city in  different contexts such as transportation~\cite{zipf1946p,jung2008gravity}, migration~\cite{park2018generalized} and most recently smart mobility such as dock-less bicycles~\cite{li2021gravity,yu2021understanding}. In this study,  we analyze large-scale longitudinal park visitation  data from a year prior to the COVID-19 pandemic and one year during. Our data is based on the aggregated mobility data from mobile phones which captures the visitation count of 3,665  parks and recreational sides of different sizes and functionalities across the  state of Washington, USA. Our analysis makes the following actionable insights for researchers  and the community of practitioners: On one hand, we show that the statistics of park visitation obey the general gravity model and are  {robust} with variations on population-related parameters (i.e., socio-economic level of visitors), spatial-related parameters (i.e., size of the city) and a temporal-related parameter (i.e., seasons).  We also show that the gravity model exhibits different parameters  {pre} and  {during} pandemic for different socio-economic groups of the population. Simply put, our results indicate  that park visitation behavior  for lower-income level class exhibited the least changes   during the pandemic, however, this change is a function of random variation and is not predictable. Furthermore, for the higher income level class we observe  more pronounced changes in park visitation behavior during the pandemic which can be explained as the shorter distance  visitations have increased at a greater rate.  Comparing parks in different spatial city groups indicates  that the visitation pattern for parks located in large cities experiences a change opposite to the one that the visitation pattern for parks located in small cities. In larger cities, the behavioral change has focused on a reduction in visitations to further distant parks, and in smaller towns behavior has changed towards an increase in further afield  visitations. 

Our analysis of modeling  park visitation  as a network structure shows that the in-degree distribution of parks is  {power-law}, both pre-and during the pandemic. Moreover, we  find a  linear association related to both  changes in visitor diversity (in-degree of parks) and the recreational  propensity of residents (out-degree of census blocks) due to the pandemic.
The higher the socio-economic level of the census block, the less change in their recreational propensity, maintaining a larger set of recreational options throughout the pandemic. Finally, from the park perspective, we observe that the larger parks exhibited a greater positive  change in the diversity of visitors, and parks that were more popular (measured as visitor density) prior to the pandemic exhibited a drop in the diversity of visitors during the pandemic. 

\section{Results}
\subsection{Gravity Model as the Robust Park Visitation Model in Washington State}\label{gravity}

We aim to study the park visitation behavior by the examples of parks in Washington State and how it changes during the COVID-19 pandemic. In particular, we want to understand how the pandemic affects the behavior of different socio-economical and spatial groups in terms of park visitation. In doing so, the first step is to find the general pattern that park visitation obeys. 
The gravity model in its general form asserts that the number of visits from the origin (i) to destination (j), noted as $T_{ij}$, is proportional to the number of visitors living in  the origin $i$, $O_i$, the attractiveness of the destination j, $D_j$, a normalization weight $W$, and finally, there is a decreasing function, $f(c)$, in terms of the cost of the visit, $c$, which is dependent on the distance between  {i} and  {j}. The gravity model in its general form can be formulated as Equation \ref{eq:gen_gravity}.
\begin{equation}\label{eq:gen_gravity}
       T_{ij}=W \;O_i\;D_j f(c)
\end{equation}
The gravity model is more general than other state-of-art visitation models, such as the radiation model~\cite{simini2012universal} and the intervening opportunities model~\cite{stouffer1940intervening} as all these models are special variants of the gravity model by choosing the proper cost function, f(c)~\cite{barbosa2018human}. Therefore, we consider park visitation obeys the gravity model and tune the parameters of a general gravity model in a way that fits the park visitation patterns. 
\begin{figure}[!ht]
    \centering
    \includegraphics[width=1\textwidth]{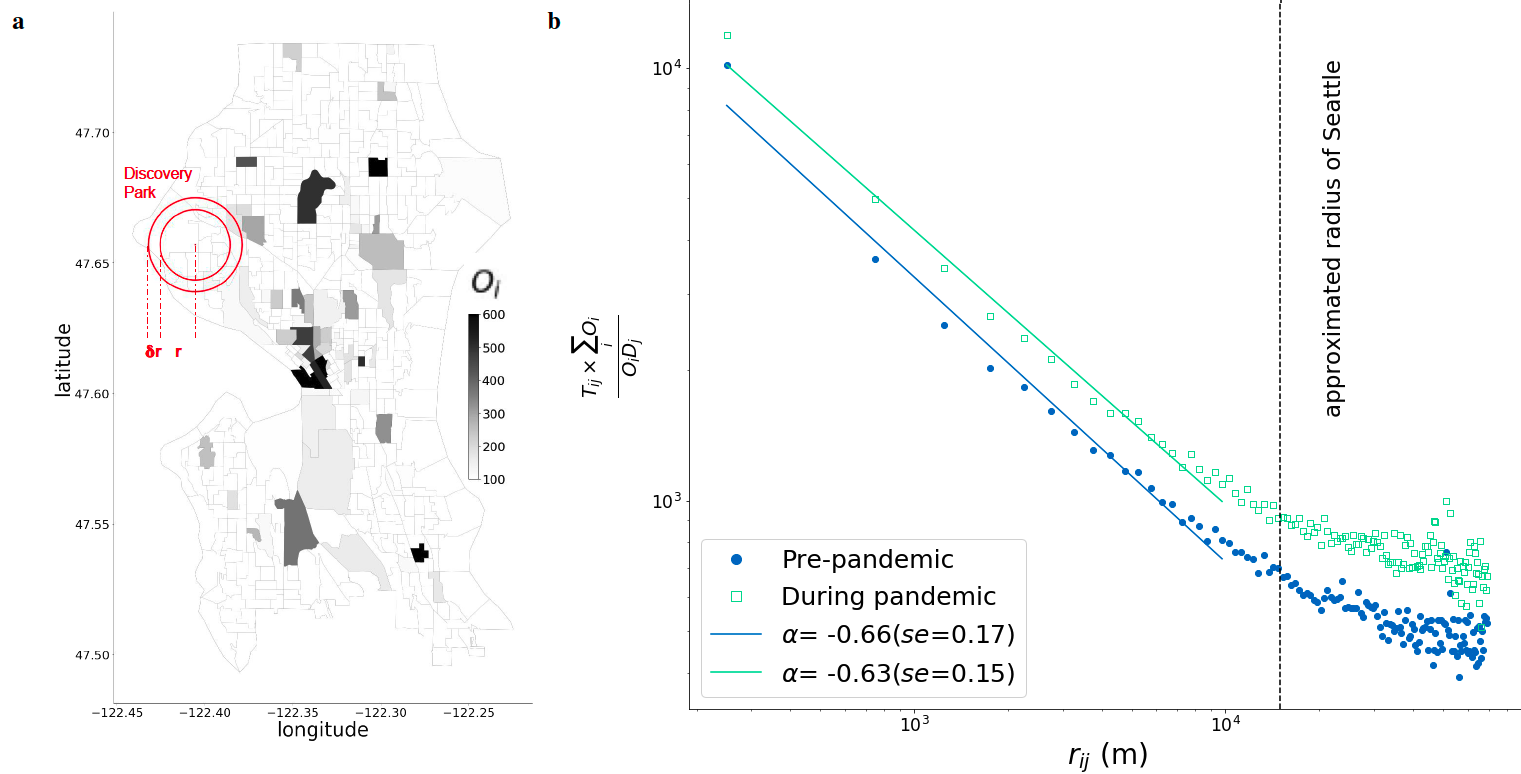}
    \caption{{\bf Gravity Model as the Robust Park Visitation Pattern}. ({\bf a}) For each park, we aggregate our data using distance bins equal to $\delta r=0.5 km$ (see details in  { Methods}). The Y and X axes indicate longitude and latitude, respectively. The map colors indicate the number of visitors residing in each census block group. {\bf (b)} The fitted (lines) gravity model plots for both pre- (blue circle marker) and during (green square marker) pandemic. The y-axis represents the visitation demand for the park for CBG, and the x-axis represents the distance between the park and CBG. The park visitation behavior obeys the gravity model both pre- and during the COVID-19 pandemic. As the parameter $\alpha$ remains unchanged, the pandemic does not affect the rate at which the visitation demand of the park changes with the distance from it. }
  \label{fig:grav_model}
\end{figure}

To explore the park visitation patterns, we consider the home census block group (CBG) of a visitor, $CBG_{i}$, as its origin, and the park it is visiting, $park_{j}$, as the destination. The home census block group for a visitor is the CBG the visitor spends most of the nighttime hours there (see  Methods). Our data is spatially aggregated, with the number of visitors originating from the $CBG_{i}$ as $O_{i}$. In addition, each $park_{j}$ has $D_{j}$ number of total visits from origins. Since the number of visits from $CBG_{i}$ to the $park_{j}$, $T_{ij}$, depends on the distance between them, we need to calculate the distance between each pair of CBG and park. The distance between $CBG_{i}$ and $park_{j}$, $r_{ij}$, is considered as the geodesic distance between the park location and the census block location. In this problem, the gravity model can be formulated as: 

\begin{equation}\label{eq:gravity}
 T_{ij}=\frac{O_{i}}{\sum_{i} O_{i}}\times\frac{K D_{j}}{ r_{ij}^{\alpha'}} 
\end{equation}

with $K$ and $\alpha$ parameters that are estimated based on the data fit.  To measure the uncertainty   of $\alpha$, we compute  the square root of the diagonal elements of the covariance matrix of coefficients, known as  the standard error of the coefficient, denoted by "se".

We used aggregate mobile data from Washington State, USA. This data covers the visits to 3665 parks from 220426 residential census blocks (See  Methods for a full description of the dataset). We separated our data into two categories pre-pandemic and during-pandemic. The pre-pandemic category starts from March 2019 upto but not including March 2020, and  the during-pandemic   category starts from March 2020 until March 2021. Additionally, the visitation data is spatially aggregated up to the size of census block groups, therefore, we use granulating bins of equal length $\delta r_{ij}$ = 0.5 km to calculate the distance (Figure \ref{fig:grav_model}-a). 
We now show that $T_{ij}$ for both above categories obeys the gravity model in the form of Equation \ref{eq:log_gravity}, where $\alpha=-\alpha'$.

\begin{equation}\label{eq:log_gravity}
    log \left( \frac{\sum_i O_i}{O_i}\times \frac{T_{ij}}{D_{j}} \right) = \alpha\; log(r_{ij})+ log(K)
\end{equation}

% \section*{Results}
Figure \ref{fig:grav_model}-b shows the gravity model for pre and during-pandemic data. In this plot, the slope of lines represents the parameter $\alpha$, and the y-intercept represents $log(K)$ (Equation \ref{eq:gravity}). Put simply, the parameter $\alpha$ quantifies the unlikelihood of  visiting a distant park. That is the greater the $\alpha$ is, 
visitation of closer parks becomes more frequent. In both categories, the park visitation obeys the gravity model accurately. Another observation is in regards to the parameter $\alpha$  (i.e., the slope of the trend) where we can see that $\alpha$  does not exhibit any significant change. The standard error ($se$) assesses the goodness of fit for $\alpha$. However, to examine the significance of  the difference  between the pre-pandemic and during-pandemic data sets, we conducted a pairwise z-test (See Methods). Throughout the rest of this article, we exclusively report differences in $\alpha$, indicating a preference for shorter or longer distances, only when the compared data sets show statistically significant differences, irrespective of the goodness of the fit. To address the challenge of large overlaps in standard errors and enhance the reliability of the interpretation, a larger sample size analysis is required.

In Figure \ref{fig:grav_model}-b, we found  that the two distributions are  significantly different, as evidenced by a z-stat  of 2.35 $(p < 0.01)$.  This observation means that although during the pandemic park visitation has significantly dropped (Supplementary Figure 10), the general trend regarding the traveled distance has remained the same as a whole.  As our goal is to study the behavioral changes  in park visitation  due to COVID-19, we dis-aggregate our data  to correspond to  different seasons, socio-economic, and spatial groups and analyze if and how selecting a subset of our original data affects the slope of gravity model ($\alpha$). We remark that we do not investigate $K$ since it is a normalization parameter so that the sum of the visitors to $park_j$ from all CBGs, $\sum_{i}T_{ij}$, is equal to $D_j$. Thus in the following sections, we will use filters based on the seasons, the socio-economical vulnerability of visitors, and the spatial scale of cities to understand the differences between various behaviors of these groups during the pandemic.

\subsubsection{Filtering by Seasonal Patterns}

As the literature on park visitation has previously shown~\cite{hewer2016seasonal,wilkins2021social}, there is a large degree of variation in park visitation trends, both the number of visits and spatial behavior of visitors, during warmer and colder seasons. To account for seasonality in park visitation trends, we divided our data into two temporal categories corresponding to the warm season when the average monthly temperature is higher than the average yearly temperature (corresponding to May to October), and cold seasons  with a mean monthly temperature lower than average yearly temperature (November to April). 

\begin{figure}[!ht]
  \centering
    \includegraphics[width=1\textwidth]{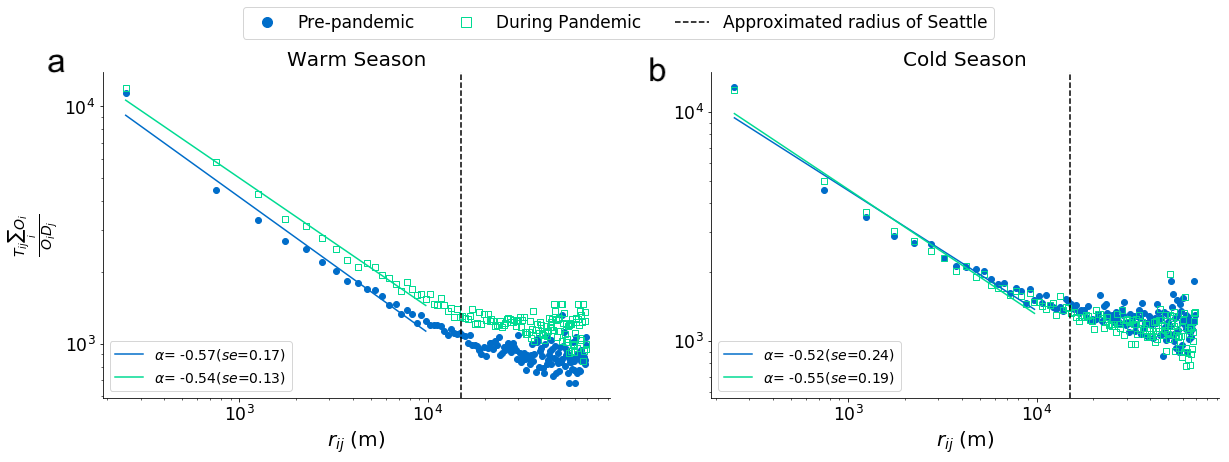}
  \caption{{\bf Gravity Model for the Seasonal Filter.} {\bf (a)} The fitted gravity model plots for warm season. The gravity model is an accurate fit for the park visitation patterns for both pre- (blue circle marker) and during-pandemic (green square marker)  and the two distributions are significantly different ($z-stat=2.56, p-value < 0.01$).{\bf (b)} The gravity model correctly portrays the park visitation patterns for the cold season.  The change in the parameter $\alpha$ due to the pandemic is  {not} {statistically significant} using solely the cold season filter. }
  \label{fig:seasonal}
\end{figure}

Figure~\ref{fig:seasonal} shows that the gravity model   accurately describes the park visitation pattern in cold and warm seasons separately. This shows that the gravity model is robust in describing the park visitation pattern after using temporal filters. However, in comparing the two distributions during the cold season using a paired sample z-test, we observe no significant difference between pre- and during the pandemic.
Furthermore, we observe no difference in parameter $\alpha$ pre- and during the pandemic after using the temporal filter. As  previously mentioned, the magnitude of park visitation is different in cold versus warm seasons, therefore we will keep the warm seasonal filter throughout studying the gravity model using socio-economic and spatial filters.

\begin{figure}[!ht]
\centering
    \includegraphics[width=1\textwidth]{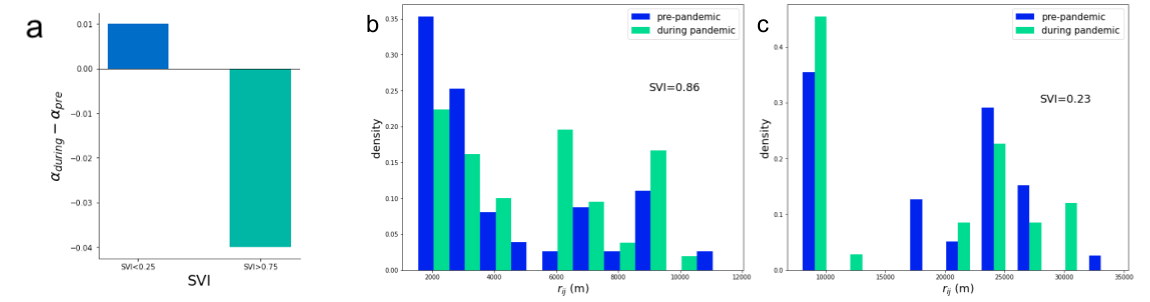}
      \caption{{\bf Gravity Model for Socio-economical Filters During Warm Season.} This figure shows the change in park visitation patterns for different socio-economic groups. {\bf (a)} Compares the effect of the pandemic in the least vulnerable  group in blue (SVI$<0.25$) with the most vulnerable group in green (SVI$>0.75$). In the least vulnerable group, $\alpha$ slightly increased, while it decreased in the most vulnerable group. This means in the least vulnerable group, where $\alpha$ increases, distant parks become unpopular to visit (i.e., the richer sub-population recreates within the parks in their vicinity). Whereas the most vulnerable group exhibits a negative change in $\alpha$, that is they are more likely (than their wealthy counterparts) to visit distant parks. {\bf (b)} Shows the distribution of distance to parks visited by a CBG with SVI$=0.86$ (The most vulnerable group), and {\bf (c)} Shows the distribution of distance to parks visited by a CBG with SVI$=0.23$ (The least vulnerable group) with blue bars presenting pre and green bars presenting during the pandemic. Supporting plots are shown in Supplementary Figure 11.}
    \label{fig:SVIs}
\end{figure}

\subsubsection{Filtering by Socio-economic Level of Visitors}
Previous studies have already shown that the decrease in mobility due to the COVID-19 pandemic is related to the socio-economic level of individuals \cite{gozzi2021estimating,wright2020poverty,fraiberger2020uncovering,weill2020social}. Besides, parks and green spaces have been disturbed unequally between different socio-economic groups \cite{wen2013spatial,mitchell2011comparison}. To indicate the difference in park visitation patterns of different socio-economic levels, we divided our data into different groups based on the  Social Vulnerability Index (SVI) of the origins (see  Methods Section) and focused our analysis on socio-economic groups that capture the most and least vulnerable parts of the society as commonly defined by sociologists\cite{deziel2023assessing}.  The first group is the group with origins of visit  from regions with  SVI$>0.75$, named as  {the most vulnerable} group, indicating the group of visitors which are the most vulnerable socio-economically speaking. The second group, which includes the group of  visits originating from regions with SVI$<0.25$ named    {the least vulnerable} group.
In order to study the effect of the COVID-19 pandemic on park visits of these groups,  let us first consider the gravity model for the warm season (shown in Supplementary Figures 11 panels c and d). 

 In the warm seasons, we can see that both the effect of the pandemic and the differences between the most versus the least vulnerable group become more visible, Figure \ref{fig:SVIs}-a. In the warm season, the parameter $\alpha$ in the least vulnerable group increases positively ($ z-stat = 3.05, p-value < 0.005$), while in the most vulnerable group, it decreases ($z-stat=3.48, p-value<0.0005$). In both cases, the two distributions were significantly different from each other (see Supplementary Figure 11).

This observation means that while using the socio-economic filter, the effect of the COVID-19 pandemic on the park visitation behavior in the warm season is greater than in the cold season. Moreover, the effect of the pandemic on the behavior of park visitors from the most socio-economic vulnerable group is the opposite of its effect on the behavior of the least vulnerable socio-economic group.
While during the pandemic distance starts to play a bigger role as the cost of a visit for the least vulnerable visitors, the most vulnerable visitors exhibit a less association between distance and propensity of park visits.  Figure~\ref{fig:SVIs} panels b and c present these observations for two different socio-economic groups.

In general, a subset of data, whose selection is not based on a simple random sampling (or sub-sampling), does not necessarily guarantee to represent the original (or same) statistical distribution. In other words, here selecting the visit of the most or the least vulnerable groups does not guarantee to fulfill the gravity model again. Nevertheless, in this section, we observed that the data from these sub-samplings still obey the gravity model, albeit with different values of parameter $\alpha$. As a plausible mechanism to justify the changes in $\alpha$, we described a basic null model (see  {Methods}). We considered two different random individuals' changes of behaviors, similar to white and pink noise, and showed how they alter $\alpha$ in the gravity model. In summary, the addition of uncorrelated randomness can lead to a change opposite to that observed when adding distance-correlated randomness.  In studying the null model we can observe the following:

 {Pre-Pandemic:} Adding  white noise to the fitted gravity model of the middle socio-economic class, defined as those originating from census blocks with 0.25<SVI<0.75, we can reproduce the parameter  $\alpha'$ of the pattern for the  {least} vulnerable groups of society. Whereas to reproduce the same distribution, (i.e. same parameters), of the most vulnerable group pre-pandemic, we needed to add  pink noise to the middle class. In other words, reaching the behavior of the most vulnerable group from the behavior of the middle class requires random changes that are correlated to the distance.

 {During the Pandemic}: We can explain the variation in park visitation between the middle class and least/most vulnerable population by the addition of  pink noise. That means that the variation between the middle class and the least/most vulnerable class in visits with longer distances is greater.
The deviation of the least vulnerable group from the middle class is at its highest during the pandemic. While the behavior of the least vulnerable group gets farther from the middle class during the pandemic, the behavior of the most vulnerable group gets closer to the middle class during the pandemic. The variation of the most vulnerable class from the middle class is at its highest rate pre-pandemic.
These observations are in agreement with the null model's results, demonstrating how the addition of white or pink noise influences the slope. Nevertheless, in order to conclude concretely, a cross-check analysis with individuals' data is needed, which due to privacy regulations we had no access to.
\subsubsection{Filtering Parks by Size of Cities}
\begin{figure}[!ht]
\centering
    \includegraphics[width=1\textwidth]{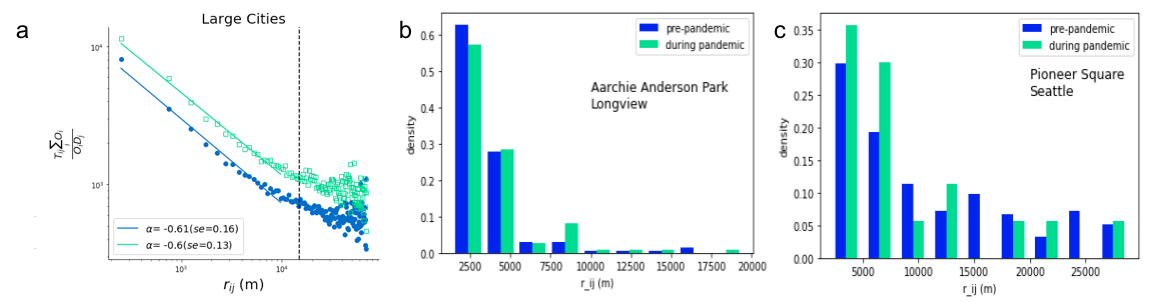}
      \caption{{\bf Gravity Model for Spatial Filters.} This figure shows the change in park visitation patterns for different spatial groups. {\bf (a)} shows the effect of the pandemic on park visitation behavior in large cities with  pre- (blue circle marker) and during (green square marker) pandemic. The parameter $\alpha$ has increased in large cities and distant parks have become less popular. On the other hand, this observation is reversed in small cities and $\alpha$ decreases significantly. {\bf (b)} shows the probability density function (PDF) illustrating  the distribution of the distance visitors have traveled to visit the Aarchie Anderson Park, located in a small city (Longview). {\bf (c)} shows the  probability density function (PDF) illustrating the distribution of the distance visitors have traveled to visit   Pioneer Square, located in a large city (Seattle). Supporting plots are shown in Supplementary Figure 12.}
    \label{fig:size}
\end{figure}

Previous research shows that people living in larger cities commute longer distances in their daily travels~\cite{kang2012intra}, and thus   spread of COVID-19 has been shown to increase with the size of the  city~\cite{stier2020covid}. For these reasons, we investigate the visitation pattern and its changes due to the pandemic for parks located in different city sizes. To do so, we divide parks into two groups based on the population of the city they are located in. In defining city boundaries we used information provided by the Bureau of the Census to map our data to cities. The first group includes parks located in the 10 most populated cities in Washington State. These ten cities,  Seattle, Spokane, Tacoma, Vancouver, Bellevue, Kent, Everett, Spokane Valley, Renton, and Kirkland, have an overall population of 49.4\% of the Washington state population. We refer to this spatial grouping as  {"large city groups"}. The second group includes parks located in the remaining Washington state cities. We refer to this group of parks as  {"small city groups"}. In studying the behavior of these two groups of parks, we include  our warm season  filter. We observe that the park visitation pattern obeys the gravity model for both spatial groups. We observe the statistical significance of  differences in pre and during-pandemic distributions for large cities ($z-score=3.88, p-value<0.0001$). However, we do not observe this difference in small cities.
Figure \ref{fig:size}-a shows that the COVID-19 pandemic had no impact on the parameter $\alpha$ of the gravity model ($\alpha$ remains approximately $-0.6$ for large cities and $-0.5$ for small cities). This difference in the $\alpha$ parameters between different types of cities means that for parks in large cities, the pandemic has led to a sharper decline in  the number of visitors, and  more significantly so with the increasing distance from parks. The number of visitors  to parks in small cities decreases slower   compared to larger cities.
Figure \ref{fig:size}-b and c illustrate the probability density function for a randomly selected park in a small city (b) and a large city  (c).  we can see that the COVID-19 pandemic has affected park visitation in both small and large cities more significantly.

Thus far we have shown that the visitation patterns differ by size of the city and by season. One way of studying what deviation from the visitation pattern in small cities develops the pattern that we see in large cities is to add noise to the visitation patterns in small cities. In the pre-pandemic period, we built the large cities' pattern by adding pink noise to the pattern for small cities (Figure 12 in Supplementary Material). However, in the case of patterns during the pandemic, we added pink noise with higher intensity to the pattern for small cities to rebuild the pattern for large cities (see Figure 12 in Supplementary Material). Referring to the null model hypothesis (see  Methods), this can be interpreted that  during the pandemic, the deviation of pattern for large cities from the small cities decreases more with the distance. This shows that the pandemic causes short-distance transportation to differ more while comparing park visitation in small vs large cities. However, as mentioned above a cross-check with individual traces is needed.

\subsection{Visitation Network}
The results presented so far have focused on the population-level analysis of park visitation data. We have shown that  the gravity model holds for  park visitation and moreover, it reveals  characteristics related to recreational  patterns before and  during the pandemic. As the gravity model is primarily based on the distance between the home location of the visitors and parks,  changes in its parameters are also a way to quantify the impact of traveled distance  on visitation patterns of different groups of people during the pandemic.

However, an alternative way of examining the impact of the pandemic on recreational  {visits} is to study the visitation patterns independent of the traveled distance and rather focus on the  structural patterns pre and during the pandemic. In other words, we seek to understand whether the visits have any predefined structure  or happen at random. To answer this question, we adapt network analysis, which represents the visits' structure, to understand and quantify the difference in the visitation network.  
 
 Network analysis has been heavily employed in understanding the diversity of regions and in relation to socio-economic status.  Eagle et al.~\cite{eagle2010network} found a strong correlation between the number of each individual's relationships and the economic development of their community. Chang et al~\cite{chang2021mobility} have modeled the spread of the COVID-19 virus using fine-grained dynamic mobility networks. They showed that  racial and socio-economic disadvantaged groups visit more crowded places and were not able to reduce their mobility as sharply due to the pandemic, thus they suffered from a higher infection rate. From a policy-making perspective, network analysis can help provide information about the resource allocation of parks and recreational areas. For example, the extent to which a park serves the local residents of the immediate area versus attracting people from all across the region is a consideration in resource planning, from creating more cycling routes and facilities at a park to planning for alternative public transportation stops and routes. Furthermore, knowing about the visit propensity of people to travel outside of their residential neighborhood can also indicate the need for addressing equity to park access across the population. For instance, Seattle Park and Recreation state their mission on increasing equity in access to parks and distributing   resources such that it  ``takes into account past history and current position, so that future outcomes are fairly distributed''~\cite{seattlerec}. In this section, we model our data as a visitation network as detailed next.
 
 \subsubsection{Inferring Visitation Network} In the field of network science, the concepts of "in-degree" and "out-degree" hold pivotal significance in elucidating network dynamics. These fundamental metrics serve to  measure   the quantification of efflux or the extent to which an entity disseminates connections/interactions to other entities within the network (i.e., out-degree) and   the extent to which an entity is the recipient of connections or interactions from other entities within the network (i.e., in-degree). Thus
park visitation network based on the movement of visitors from CBGs to parks can be presented as a directed bipartite network in the form $G=(U,V,E)$, where: $U=\{u_1,u_2,\dots,u_n\}$ represents $n$ CBGs,  $V=\{v_1,v_2,\dots,v_m\}$ represents $m$ parks,  $\text{E }\subseteq \{(u_i,v_j)\mid u_i\in U\text{ and }v_j\in V\}$.

We used the same SafeGraph data to build the park visitation network but we removed seasonal and socio-economic filters. In doing so, we connect the $CBG_i$ to $park_j$ only if $T_{ij}\geq 2$ (see  Methods). In making the visitation networks, park visitation has been considered from March 2019 till March 2020, a year before the start of the pandemic (pre-pandemic network), and again from March 2020 until March 2021, a year after the start of the pandemic (during the pandemic network). Each CBG is labeled with an SVI of the residential area, the number of residing devices, and its area in  square kilometers. Each park is labeled with an SVI related to its location, the total number of visitors, and its area   in square kilometers.   

\begin{figure}[!ht]
  \centering
    \includegraphics[width=\textwidth]{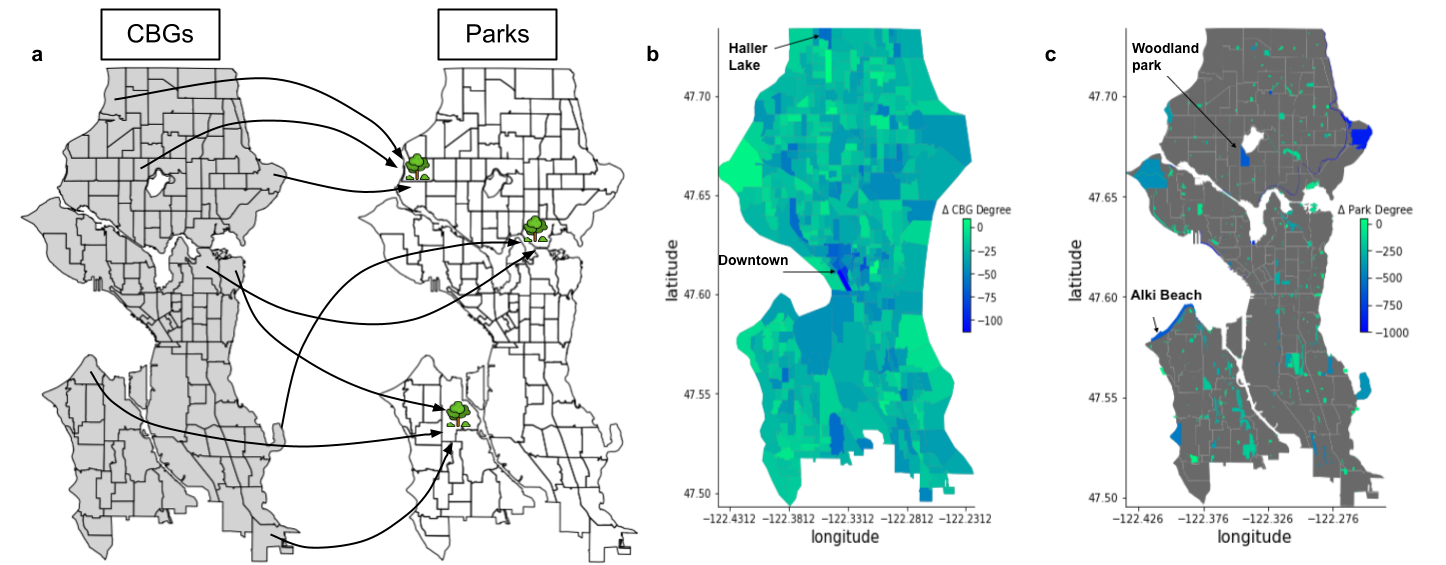}
  \caption{{\bf The Park Visitation Network.} {\bf (a),} A schematic representation of how the  Park Visitation Network is built up. This network is bipartite including two separated parts: CBGs and Parks which are connected by direct edges from CBGs to Parks. The degree of a park in this network is the  {diversity} of the park and the degree of a CBG is the  {propensity}. {\bf (b),} The map shows CBGs in Seattle, and the color map indicates the change in  {propensity} of CBGs due to the COVID-19 pandemic. The pandemic has decreased the  {propensity} of CBGs of Seattle. {\bf (c),} The map shows parks in Seattle, and the color map indicated the change in  {diversity} of parks caused by the COVID-19 pandemic. As shown in the map, the pandemic decreases  {diversity} of parks located in Seattle. The time windows of the pre-and during the pandemic are similar to Figure \ref{fig:grav_model}. Data is not filtered seasonally or by any socio-economical group.}
  \label{fig:network}
\end{figure}

% \section*{Results}
Figure \ref{fig:network}-a    illustrates the schematic presentation of park visitation as a bipartite network in which each CBG is connected to multiple parks with unweighted edges. Each link between the $CBG_{i}$ and $park_{j}$ means that there are at least two visitors who live in $CBG_{i}$ and visited $park_{j}$. Thus the network analysis focuses on two sides: firstly, the CBGs' out-degree (Figure \ref{fig:network}-b) which presents the  {recreational propensity} of the CBG residents going to parks. Secondly, the parks' in-degree (Figure \ref{fig:network}-c), which presents the  {diversity} of park visitors originating from different census blocks.  

As presented in the heatmaps both diversity and recreational propensity have decreased during the COVID-9 pandemic. Regarding the propensity (Figure~\ref{fig:network}-b), we can see that some areas (such as Haller Lake and Downtown) exhibit a sharp decline in the magnitude of visits to parks outside of their local area (out-degree edges). As we show next this decline in CBG degree is higher for low-income and highly populated areas in the city. We also observe a great reduction in the diversity of the visitors (incoming edges) to the parks (Figure~\ref{fig:network}-c). As our analysis will show parks with higher demand face a greater degree of park degree changes. Some of these changes could also be due to other major factors besides the pandemic. For example, the closure of the West Seattle Bridge during 2021-2022 is known to have reduced accessibility to Alki Beach (Figure~\ref{fig:network}-c), and Woodland Park closure due to homeless encampment.

\subsubsection {Quantifying Network Properties}
\begin{figure}[!ht]
    \centering
\includegraphics[width=\linewidth]{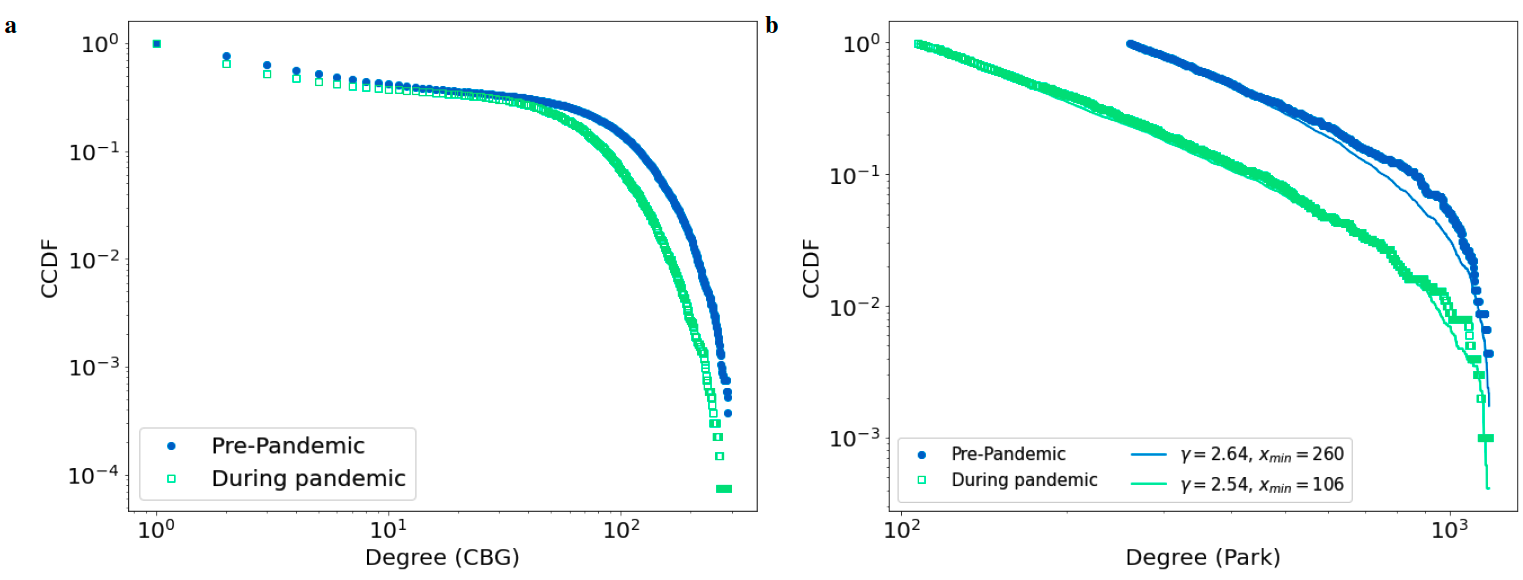}

    \caption{{\bf Cumulative distribution function (CDF) of degree for park visitation network located in Washington state,} shown as  blue circle marker for pre and  green square marker for during pandemic. It shows how the distribution of degrees for {\bf (a)} CBGs visiting parks (i.e., out-degree)   {\bf (b)} parks (i.e., in degree) changed due to the pandemic. The right panel is well fitted to a power law function:  $P(x)=\frac{\gamma-1}{x_{min}}(\frac{x}{x_{min}})^{-\gamma}$ with renormalization factors of $x_{min}$. Points show the empirical data and curves present the fits.}
  \label{fig:deg_dist}
\end{figure}

\begin{figure}[!ht]
    \centering
    \includegraphics[width=\linewidth]{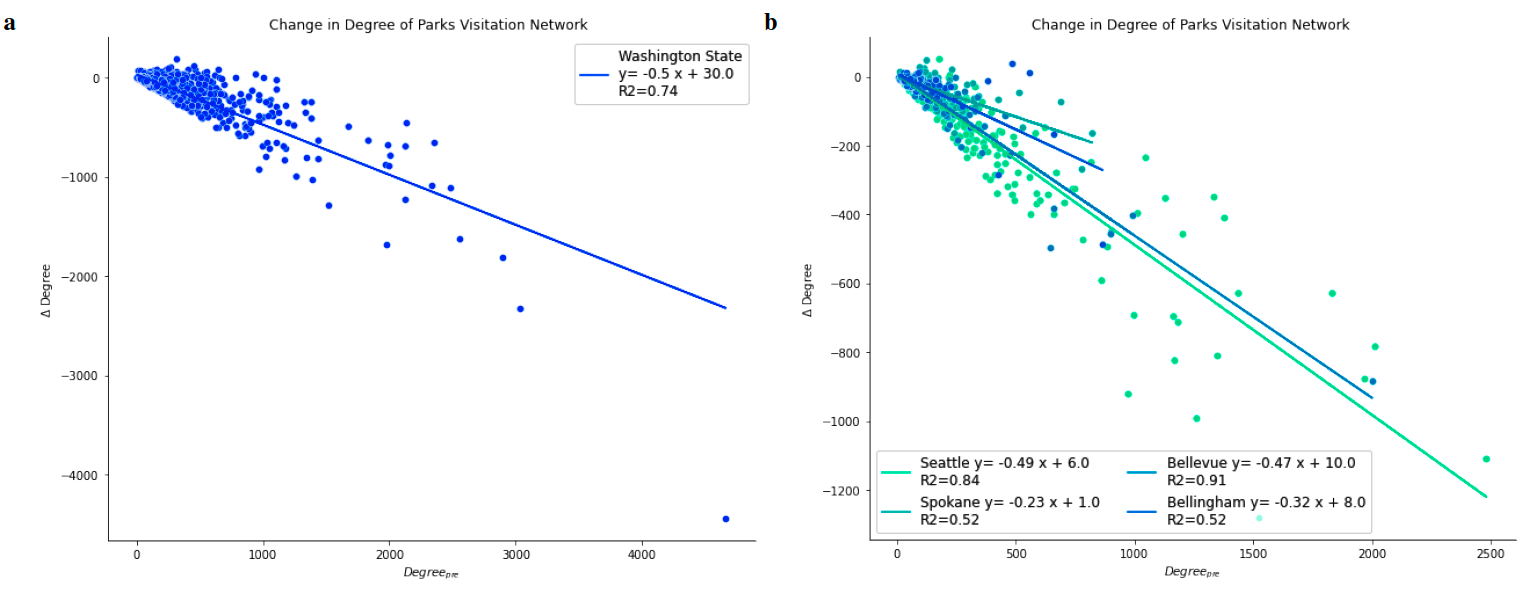}
    \caption{{\bf Relation of change in park degree with the degree of park pre-pandemic.} {\bf (a),} Plot shows the relation between the degree of parks in Washington state in the pre-pandemic visitation network and change in their degree due to the Covid-19 pandemic. Change in the degree of a park has a linear relation with the pre-pandemic degree of the park where the slope of this linear relation is equal to $-0.5$. By separating parks located in each city, we can still see this linear relationship. {\bf (b)} shows the change in park degree versus the pre-pandemic degree for parks located in the four Washington cities with the most number of parks, separately. This relation in each city is still linear, with different slopes in each city.}
  \label{fig:linear_deg}
\end{figure}

Let's continue with a quantitative analysis of this bipartite network. We first show the Cumulative distribution function (CDF) of the CBG's out-degree and also the park's in-degree (Figure \ref{fig:deg_dist}) and then the changes in the park's in-degree (Figure \ref{fig:linear_deg}).

Initially, we look at Figure \ref{fig:deg_dist}, the out-degree (panel a) and in-degree (panel b) distributions of the park visitation network, which are heavy-tailed distributions. We observe that out-degree distribution can be described neither with normal nor scale-free distributions. Nevertheless, by having renormalization factors, the power-law distributions fit the in-degree distributions as shown in Figure \ref{fig:deg_dist}-b, but the parameters of the fitted curve, $\gamma$ and $x_{min}$, change due to the pandemic. To identify the most suitable descriptive law for these distributions, additional studies are required.

Moreover, in order to see how park degrees relatively changed, we plotted the change in the degree of parks versus the degree of the park in the pre-pandemic network for all the parks in Washington State (Figure \ref{fig:linear_deg}-a) and four large cities in Washington state separately (Figure \ref{fig:linear_deg}-b). Figure \ref{fig:linear_deg} shows that the degree of parks changes linearly with their degree in the pre-pandemic situation with a negative slope. This means that parks with visitors from a wider range of CBGs experience a bigger decrease in the number of CBGs that visit them. Since the pandemic causes a decrease in the CBG degree (Figure ~\ref{fig:deg_dist}-a), therefore it's reflected in  {propensity}, as well (Figure \ref{fig:network} maps this decrease for Seattle CBGs). 

\subsubsection {Network Correlation Analysis}
Given the park visitation network, now we want to know whether  {diversity} of parks and  {propensity} of CBGs depend on socio-economic status and geometric scale. This question can be answered by measuring the linear correlation between degrees in the visitation network and SVI, Area (park or CBG), and number density (visitors/ residents). The main observations of this analysis are the following: 

First examining propensity (Table 1), we observe a strong statistically significant correlation between the CBG degree and SVI. That is the higher the social vulnerability of an area, the residents of that area exhibit a lower propensity in park visitation, both pre ($\rho= -0.93$) and during the pandemic ($\rho= -0.90$). Therefore, residents of the least vulnerable areas (i.e., low SVI) visit a broader range of parks compared to the most vulnerable areas (i.e., high SVI). However, there is a high positive correlation between the change in the CBG degree and SVI ($\rho=0.87$). This shows that if the SVI is higher (more vulnerable), then the change in the degree due to the pandemic is also higher. Therefore, the pandemic has a higher effect on the propensity of the most vulnerable CBGs compared to the least vulnerable ones. We also observe that the population density (as measured by devices per area) for different census areas exhibits a negative correlation with the change in park propensity ($\rho=-0.70$). Simply put, residents in the denser areas (such as the Downtown area in Figure~\ref{fig:network}-b consisting of apartment blocks) reduced their choices of recreational visits. We remark that this observation could be specific to characteristics of urban design and lifestyle in the United States and more specifically to Washington state where socioeconomic status and population density are intertwined. 

Second examining diversity (Table~2), indicates contrasting observations regarding the SVI. That is  parks in richer areas of Washington State did not exhibit a significant difference in their visitors' diversity. However, we observe that parks located in higher-density areas have a higher diversity  during the pandemic ($\rho=0.94$). Moreover, the change caused by the COVID-19 pandemic in the degree of larger parks (area) is higher compared to smaller ones ($\rho=0.79$). This shows  that the increase in diversity of larger parks is higher due to the pandemic.  Our results show that the park area does not have a strong correlation with the degree of the park pre and during the pandemic, the change in diversity is highly correlated with the park area. We also show that the CBG degree is highly correlated with SVI both pre and during the pandemic. This also results in a correlation between the change in CBG degree and the SVI. 

\section{Conclusions}
 In summary, we have analyzed the park visits of Washington State pre and during the pandemic from two perspectives: on one side, we showed that visits obey scale-free laws, namely, the gravity model and this pattern is robust, even though the slopes change due to different factors. On the other side, we inferred a  network of visits, measured some network properties, and showed how the propensity and diversity of the visits were changed and which factors could play a role in it. We also showed that the in-degree and the out-degree of recreational visits from neighborhoods are heavy-tailed distributions. To identify the most suitable descriptive law for these distributions, additional studies are required. Overall, our results shine a light on  {practical implications} for policymakers and park managers on one hand and raise some  {theoretical questions} on the other hand.
 
 Putting all the reported observations together, our analysis highlights the following practical implications: due to the pandemic the higher-income level residents changed their behavior by visiting more local parks and broader recreational options outside of their local census area; whereas the low-income residents changed their visitation behavior greatly by reducing their recreational choices outside of their local area and their traveled distance did not follow a predictable pattern. These behavior changes might be due to the lack of access to specific facilities in parks or the closure of parks in the most vulnerable areas. It could also resonate with the greater  work-life balance imposed by the pandemic for only a certain group of the population (i.e., economically resilient). A practical implication of this result is that policymakers can potentially plan alternative recreational activities in areas with low socio-economic levels. Neighborhood programs such as   meet-ups, outdoor clubs, etc. that take  steps to prevent transmission of the virus but  target specifically low-income residents  to be more active could help bring equitable recreational access. 

We also observed that  the recreational behavior  in small versus large cities is   different, and thus the scale of the city is also another factor that should be accounted for in decision-making. We believe, having access to a more detailed and richer dataset could  help to understand the underlying causes of the behavioral shifts we observed in this study. For instance, rich content from social media posts along with qualitative surveys of the park visitors could help us understand more about their decisions of selecting a particular recreational site over other alternatives. Finally, having access to large-scale fine-grain (non-aggregated)  mobility data could help us to portray a more accurate picture of recreational choices.  We believe in much the same way that technology companies, scientists, and local authorities worked together to create infrastructure for tracking and tracing the spread of the COVID-19 virus, a unified effort that addresses privacy concerns upfront could be mounted to enable advanced modeling of the  recreational choices at the individual level.
 
 Our work has various theoretical implications and prospects for the research community: First and foremost our study shows that even by only  sub-sampling  the urban mobility data, namely visits to parks, the gravity model was still well fulfilled. Furthermore, additional  sub-sampling, namely three different filters of seasonality, socio-economic, and spatial ones showed that the gravity model is robust while only the slope differs for different filters and also pre and during pandemics. 

 In closing this work creates opportunities to focus  {future} studies on the following:  {i)} confirming the universality of the gravity model when applied to  recreational patterns of other cities and countries  from different ranges of GDP and cultures.  {ii)} analyzing under which conditions sub-sampling can keep the original scaling law.  {iii)} investigating how to infer the original general law by knowing the sub-sampled statistical law, for instance by knowing the slope of gravity model for a given filtered data.  { {iv)} how to improve null models in order to understand the underlying mechanisms of observed changes.} 
 
 Finally, our analysis revealed that the degree of parks in the visitation network follows a power-law distribution while the park visits obey the gravity model. Knowing these two perspectives, a research question   arises as to whether ensemble models generate synthetic park visitation traces. In the same vein, we believe future work will include applying advances of graph-based neural networks~\cite{zhou2020graph} to enhance the gravity model as a non-linear model   similar to recent efforts by~\cite{simini2021deep} to generate synthetic traces. Such models can be then adjusted for socio-economic variations using the fairness regularizer layer in their architecture to force the model to tune for different socio-economic levels during  the training process~\cite{yan2020fairness}.

\renewcommand{\arraystretch}{1.5}
\begin{center}
\begin{table}[h!]
\begin{tabular}{ l  || c  | c  | c  }
   CBG degree & pre-pandemic degree & during pandemic degree& change in degree\\ 
  \hline
  \hline
  residence SVI &  {-0.93}****&  {-0.90}**** &  {0.87}**** \\ 
  \hline
  residence area & -0.19* & 0.07* & 0.18*\\
  \hline
  $\frac{\text{\# devices}}{\text{residence area}}$  & 0.36* & 0.51* &  {-0.70}***
\label{table:CBGcorr}
\caption{Propensity Correlation. Correlation of census block group degree with residence SVI, residence area, and $\frac{\text{\# devices}}{\text{residence area}}$ for CBGs. Number of * represents p-value: * p-value<1 , ** p-value<0.05, ***p-value<0.005, **** p-value<0.0005. Here, the change in degree represents the change in the degree of the CBG due to the pandemic.}
\end{tabular}
\end{table}
\end{center}

\begin{center}
\begin{table}[h!]
\begin{tabular}{  l || c |  c| c}
   Park degree & pre-pandemic degree & during pandemic degree& change in degree\\ 
  \hline
  \hline
  park SVI & -0.14* & -0.31* & -0.04* \\ 
  \hline
  park area & 0.5* & 0.60* &  {0.79}**\\
  \hline
  $\frac{\text{\# visitors}}{\text{park area}}$  &  {0.67}** &  {0.94}*** & -0.72*\\

\end{tabular}
\label{table:parkcorrtable2}
\caption{Diversity Correlation. Correlation of park degree with park SVI, park area, and $\frac{\text{\# visitors}}{\text{park area}}$ for parks. Number of * represents p-value: * p-value<1 , ** p-value<0.05, ***p-value<0.005, **** p-value<0.0005. Here, the change in degree represents the change in the degree of the park due to the pandemic. In addition, the \# visitors in $\frac{\text{\# visitors}}{\text{park area}}$ has been considered to be the number of visitors pre-pandemic while comparing with the pre-pandemic degree, number of visitors during the pandemic in the second column, and the number of visitors pre-pandemic (normal situation) in the third column.}
\end{table}
\end{center}

\newpage
 \section{Methods}\label{methods}

\subsection{Data}\label{sec:data}
In this study, we used three different types of data described as follows:
 
\paragraph{ {SafeGraph Mobility Data.}}
SafeGraph is a commercial company that provides Point of Interest (POI) and Location-Based Services (LBS) data in the U.S, Canada, and the United Kingdom~\cite{safegraph} 
The dataset is an aggregated dataset of anonymized location data from numerous mobile applications.  SafeGraph captures the movement of people between POIs that have been marked as parks and green spaces in the SafeGraph Core Places API. Unlike location-based social networks, which require active participation on the part of the user to upload and share images, SafeGraph collects its data from users who have installed one of the many affiliated mobile applications on their devices. Users' locations are recorded even while they are not actively using an application. It is therefore non-participatory and platform-independent. For these reasons, SafeGraph is less sensitive to biases that would result from changes in the popularity of the use of certain applications over time~\cite{chang2021mobility,gao2020mapping,kang2020multiscale}. Furthermore, the validity and viability of SafeGraph data as a proxy of cite visitation in US national parks has been reported recently by \cite{liang2022assessing}. 
\begin{figure}[!ht]
    \centering
    \label{main:aaaa}\includegraphics[width=\linewidth]{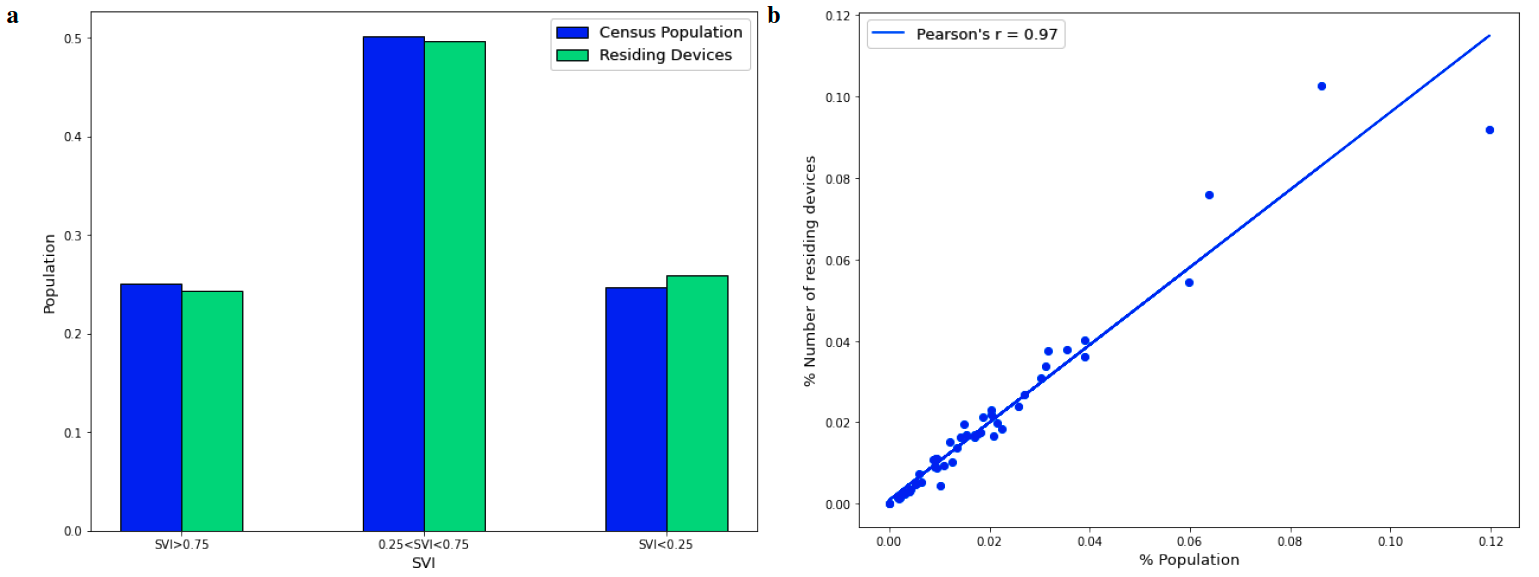}

    \caption{{\bf SafeGraph data description.} {\bf (a)} compares the distribution of SafeGraph and the U.S. census data between different socio-economic groups. The distribution of data from SafeGraph between different socio-economic groups is accurately similar to this distribution for the U.S. census data. Moreover, SafeGraph data is equally disturbed between the least and the most vulnerable socio-economic groups. {\bf (b)} Shows the correlation between the proportion of SafeGraph data in each state and the actual population of that state, estimated by the U.S census. The strong correlation shows that the SafeGraph data is spatially well disturbed.}
  \label{fig: safegraph_description}
\end{figure}
To preserve the privacy of users, SafeGraph data is provided in monthly and spatial aggregates, presenting the number of visitors and their visit characteristics such as the aggregated total number of visitors who originated from each census block group (CBG). CBGs are geographical units that typically contain a population of between 600 and 3,000 people. This rich aggregated information allows us to analyze changes in visitation patterns and visitor characteristics independent and de-coupled from online behavior. Figure \ref{fig: safegraph_description}-b shows the correlation of population with the number of devices SafeGraph has collected data from for each state. We can see that sampling from the population is correlated with the population in each state with Pearson's r equal to 0.97.  

We collected all the data associated with Parks and Green Natural areas of Washington state under a research license provided by SafeGraph. We used SafeGraph weekly pattern data from November 2018 till February 2021 for 3665 parks in Washington state and 220426 census block groups (U.S.). The population of devices in each CBG $i$, $O_i$, is given as number\_devices\_residing in this data. SafeGraph defines the home of visitors as the CBG that the visitor has spent most of their time during nighttime hours (between 6 pm and 7 am) in a 6-week period. The number of visitors for each park j, $D_j$, is given by raw\_visit\_counts, and the number of visitors from CBG $i$ to the park j, $T_{ij}$ is given by the visitor\_home\_cbgs in SafeGraph weekly pattern data. Note that SafeGraph reports the $T_{ij}$ between an origin and a destination, if the $T_{ij}$ at least is equal to two, and if $2\leq T_{ij}<4$, SafeGraph reports $T_{ij}$ as four. 

We applied the following steps in processing Safegraph data for the Gravity model.   {i)} we aggregated distances between origins and destinations using $\delta r=500$ meter bins.  {ii)} In filtering the seasonal and yearly patterns using monthly patterns, we only kept parks whose data was available for each month in a desired season/year. 

\paragraph{ {Social Vulnerability Index Data.}}
The Social Vulnerability Index (SVI) developed by the Centers for Disease Control and Prevention is a metric that captures the resilience of each community at the census tract level. This index combines 15 U.S. census variables grouped into four themes: socioeconomic status, household composition, race/ethnicity/language, and housing/transportation,  in order to rank census tracts by their relative vulnerability to hazardous events. Each of these themes has its own percentile ranking where  greater percentile values represent greater vulnerability. In this study, we used the first theme that corresponds to {\em socioeconomic} status   to understand the shift in visitors' socio-economic background pre and during the pandemic. Figure \ref{fig: safegraph_description}-a shows the distribution of different socio-economic groups in the Washington population and in the sample from SafeGraph. As shown in this bar plot, the distribution in the Washington population is the same as the distribution in the SafeGraph sampling. 

\paragraph{ {Census Data.}}
We obtained US population data from the 2019 American Community Survey (ACS) product of the U.S. Census Bureau, provided by SafeGraph’s Open Census Data. We computed a CBG’s population density by dividing the 2019 population estimate by the land area of the CBG as reported by CBG geographic data of SafeGraph’s Open Census Data. SafeGraph "uses the cartographic boundary files, which define simplified shapes of geographic entities designed for plotting, provided by the  U.S. census."

\subsection{Reporting Z-test Statistics}

To examine the difference  between two  distributions we employed a  parametric test, z-test, to determine if the means of two distributions differ. The z-test is a parametric statistical test used when the sample size is sufficiently large (greater than 30). In our case, we applied the paired sample z-test to compare the points in the gravity  model of the pre-pandemic and  during-pandemic distributions.

\paragraph{Null Hypothesis (\(H_0\))}

The null hypothesis posits that there is no significant difference between the means of the two distributions.

\paragraph{Procedure}

To test the null hypothesis:
\begin{enumerate}
    \item We calculated the mean (\(\bar{x}\)) and standard deviation (\(\sigma\)) of the data for both groups.
    \item We computed the z-score using the formula: 
    \[
    z = \frac{\bar{x}_1 - \bar{x}_2}{\sqrt{\frac{s_1^2}{n_1} + \frac{s_2^2}{n_2}}}
    \]
    where \(\bar{x}_1\) and \(\bar{x}_2\) are the means, \(s_1\) and \(s_2\) are the standard deviations, and \(n_1\) and \(n_2\) are the sample sizes of pre and during-pandemic groups, respectively.
    \item We then checked the z-score against the critical value for the chosen significance level to determine statistical significance. We set the significance level at \(\alpha = 0.05\) for all statistical tests.
\end{enumerate}

\subsection{White and Pink Noise (Null Model)}\label{sec:noise}
\begin{figure}[!ht]
    \centering
    \label{main:awn}\includegraphics[width=\linewidth]{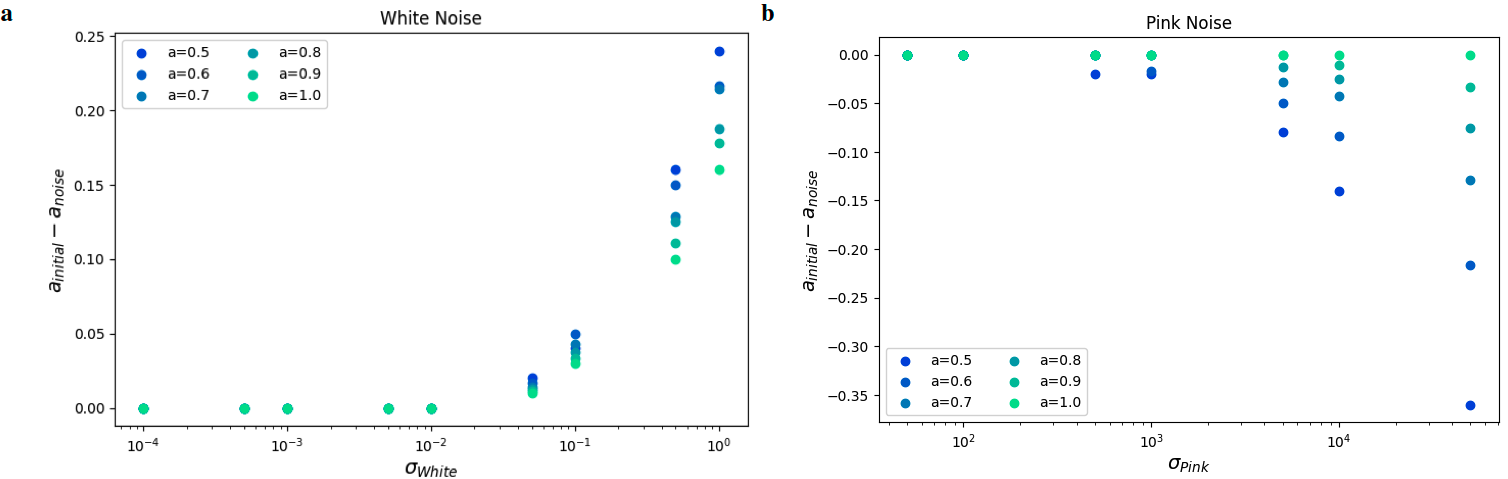}
    \caption{{\bf Adding noise to simulated data of the gravity model.} We produced a dataset matching the gravity model and measured the changes in the parameter {'a' (corresponding to the parameter $\alpha'$ in empirical data)} of the gravity model by introducing various amounts of noise to the generated data. In {\bf(a)}, the addition of white noise to the {simulated} data decreases the parameter `a,' while in {\bf (b)}, adding pink noise to the gravity model increases `a.'}
  \label{fig: pink_white}
\end{figure}

In order to understand better the changes of slops in the gravity model, let's go one step further with a simple null model. Thus we assume that the individuals' change behaviors have been random and see how the slop would change, as shown in Figure. \ref{fig: pink_white}. Simulating a fixed gravity model with  $a \in \{0.5,0.6,0.7,0.8,0.9,1.0\}$ (corresponding to the parameter $\alpha'$ in empirical data), we add two types of noises to the flow, $T_{ij}$ in the model: namely white and pink.

Firstly we consider white noise with the simplest representation of it. The term ``white'' refers to independent and identically distributed random variables that are added to our model as follows: 
\begin{equation}
    \Delta T_{ij}=\sigma_{noise}\times rand\{0,1\}
\end{equation}

where $\sigma_{noise}$ is the noise level. 

Secondly, we assume even though the change of behaviors has been random but was correlated to the distance. Thus we can consider bigger fluctuations for shorter distances which obey a power-law function.  We refer to this noise as ``pink'' noise. The noise is  produced as follows: 

\begin{equation}
    \Delta T_{ij}=\sigma_{noise}\times\frac{rand\{0,1\}}{r_{ij}+1}
\end{equation}

where $\sigma_{noise}$ is the noise level and $r_{ij}$ id the distance between CBG and a park. 

Figure \ref{fig: pink_white} shows that regardless of the initial slope, white noise (panel a) decreases $a$, while pink noise (panel b) increases the slope. Also the smaller the $a$ is or the bigger the noise level, $\sigma_{noise}$, is, the bigger the change of the slop would be.
We remark that our focus here is not on explaining why these scaling patterns are observed, as reference \cite{ribeiro2023mathematical} has already reviewed ``the main theoretical models in the literature aimed at explaining the origin and emergence of urban scaling.'' Instead, our investigation delves into possible statistical scenarios behind these changes. Therefore, we proposed the simple null model involving the addition of noise (both correlated, pink noise, and uncorrelated, white noise, to the distance) for comparison with the empirical data.

To gain a deeper understanding of the underlying processes that generate data consistent with the observed patterns, as well as the changes that occur prior to and during a pandemic, one can take a further step by utilizing null models similar to the ones presented in reference \cite{altmann2020spatial}. Furthermore, this work involves a comparison of different models in urban scaling laws, including gravity models. A similar approach can be a topic for further studies as well. To seek further elaboration on how and why the changes occurred, a cross-check analysis with individuals' data is needed. Unfortunately, due to privacy regulations, we have no access to such data.
Moreover, one can  employ a similar approach and directly add noise to a specifically empirical data set, using it as the baseline, aiming to determine the type and level of noise that could replicate a similar trend in the targeted data set. Through this exploration of noise, statistically, we can interpret that changes might have occurred randomly and could be correlated (pink) or uncorrelated (white) with distance.

\section*{Acknowledgement}
F. G. would like to thank Eduardo Goldani Altmann for his valuable comment regarding the statistical laws and models.

\bibliography{references}

\newpage
\section*{Supplementary Information} \label{appendix}

\begin{figure}[!ht]
    \centering
    \label{main:agc}
    \includegraphics[width=\linewidth]{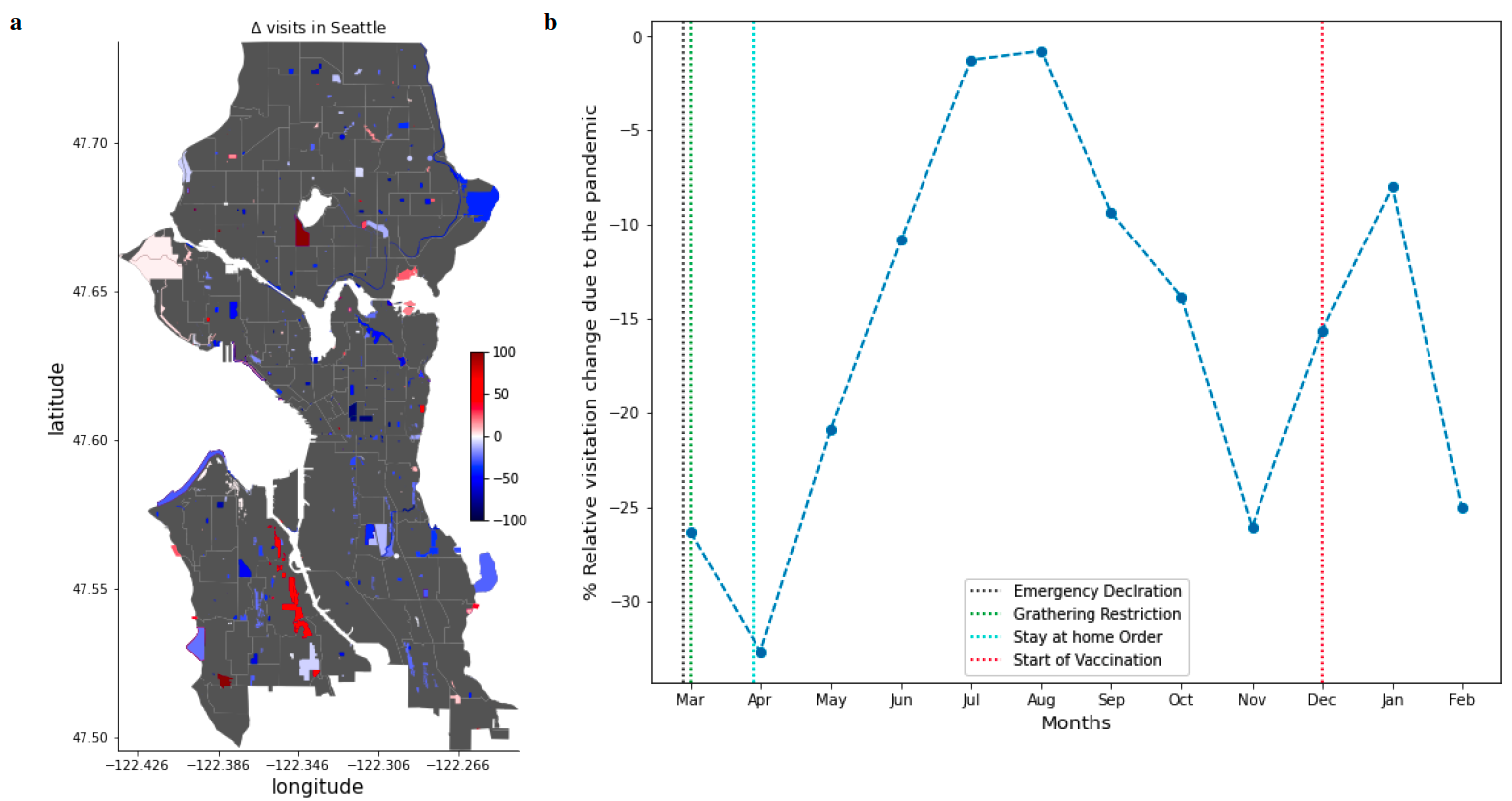}
\caption{{\bf Relative park visitation change due to the pandemic for each month, from March 2019 until February 2021} \textbf{(a)} shows the relative visitation change for the city of Seattle, demonstrating a drop in most parks, \textbf{(b)}  shows the visitation changes in relation with the statewide declaration of emergency.}
  \label{fig:general_changes}
\end{figure}

\begin{figure}[!ht]
  \centering
    \includegraphics[width=\textwidth]{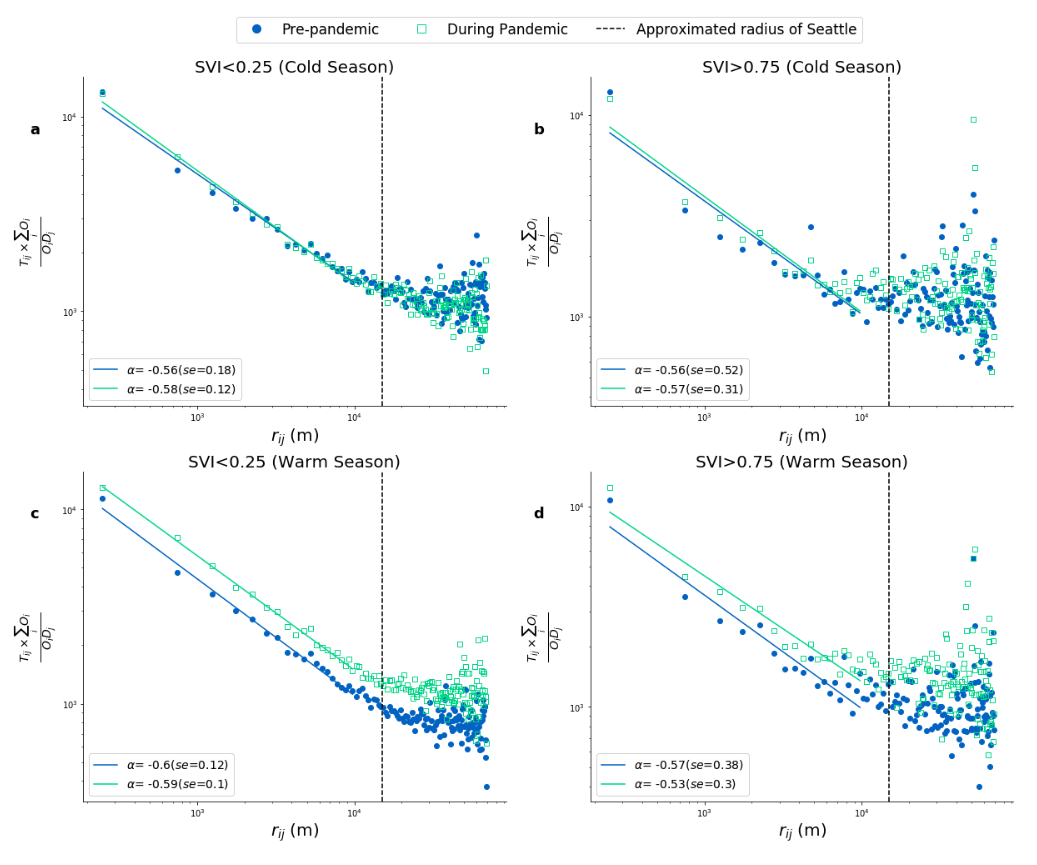}
  \caption{{\bf Gravity model using socio-economic filters.}  {Plots indicate the change in park visitation patterns for different socio-economic groups and compare the effect of the pandemic in the least vulnerable  group (SVI$<0.25$, panels a and d) with the most vulnerable group (SVI$>0.75$, panels b and c) for cold (top panels) and warm (bottom panels) seasons.   Pre-pandemic is denoted by blue circle markers  and during-pandemic by  green square markers. We observe no statistical significance in differences between pre-and during-pandemic distributions during the  {cold} season.  In the least vulnerable group (panel c), we observe less standard error  ($se$) for  $\alpha$ and a statistically significant difference between during and pre-pandemic ($z-stat=3.05, p-value<0.005$). We observe $\alpha_{during}>\alpha_{pre}$ for the most vulnerable group (panel d) with a higher $se$, with a strong significant difference between the pre and during-pandemic distributions ($z-stat=3.48, p-value<0.0005$).}
  }
  \label{fig:grav_supl_svi}
\end{figure}

\begin{figure}[!ht]
  \centering
    \includegraphics[width=\textwidth]{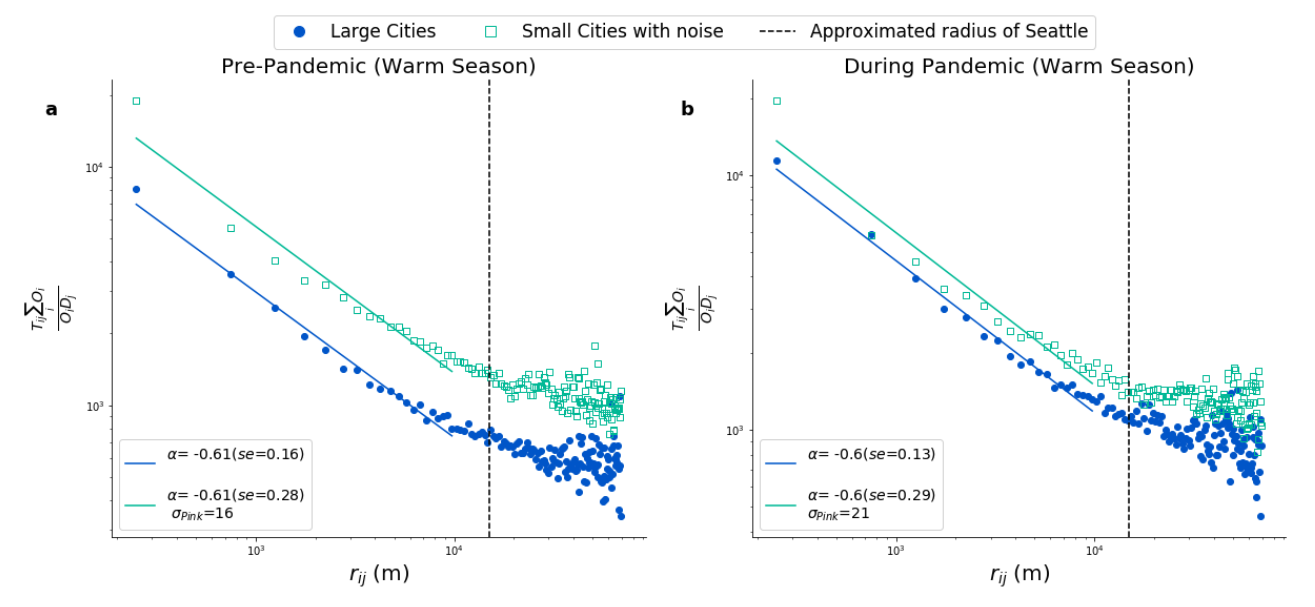}
  \caption{{\bf Gravity model for spatial filters and noise.} Supporting plots  shows the changes in $\alpha$ by adding {different intensity pink noise to the  pre-pandemic panel \textbf{(a)} and during-pandemic panel \textbf{(b)}  data. This result reflects that the deviation is dependent on the distance and to a greater extent during-pandemic.}  Large cities trend is denoted by blue circle markers  and small cities trend with added  noise by  green square markers.}
  \label{fig:city_supl_noise}
\end{figure}

\end{document}